# MFCC-GAN Codec: A New AI-based Audio Coding

**ABSTRACT**

In this paper, we proposed AI-based audio coding using MFCC features in an adversarial setting. We combined a conventional encoder with an adversarial learning decoder to better reconstruct the original waveform. Since GAN gives implicit density estimation, therefore, such models are less prone to overfitting. We compared our work with five well-known codecs namely AAC, AC3, Opus, Vorbis, and Speex, performing on bitrates from 2kbps to 128kbps.

MFCCGAN_36k achieved the state-of-art result in terms of SNR despite a lower bitrate in comparison to AC3_128k, AAC_112k, Vorbis_48k, Opus_48k, and Speex_48K. On the other hand, MFCCGAN_13k also achieved high SNR=27 which is equal to that of AC3_128k, and AAC_112k while having a significantly lower bitrate (13 kbps). MFCCGAN_36k achieved higher NISQA-MOS results compared to AAC_48k while having a 20% lower bitrate. Furthermore, MFCCGAN_13k obtained NISQA-MOS=3.9 which is much higher than AAC_24k, AAC_32k, AC3_32k, and AAC_48k. For future work, we finally suggest adopting loss functions optimising intelligibility and perceptual metrics in the MFCCGAN structure to improve quality and intelligibility simultaneously.

## INTRODUCTION

Speech coding is an integral part of any communication system. Coding helps such systems utilise the bandwidth optimally and create more capacity to deliver information. Whereas conventional coding approaches lie more on algorithmic processes, deep learning approaches are data-driven methods. Kuyk et al. compute the true information rate of speech to be less than 100 bps [2]. High-quality speech codecs typically operate at bit rates over 16 kbps [3]. Opus [4] can handle a wide range of audio applications, including Voice over IP, videoconferencing, in-game chat, and even remote live music performances. Advanced Audio Coding (AAC) [5] is an audio coding standard for lossy digital audio compression. Designed to be the successor of the MP3 format, AAC generally achieves higher sound quality than MP3 encoders at the same bit rate. It can scale from low-bitrate narrowband speech to very high-quality stereo music. Dolby AC-3 [6] is the name for what has now become a family of audio compression technologies developed by Dolby Laboratories. AC-3 supports audio sample rates up to 48 kHz. It is worth noting that AAC and AC3 are well-known broadcasting audio codecs currently standardised and utilised in different broadcasting unions.

Vorbis is a general-purpose compressed audio format for mid to high-quality audio and music at fixed and variable bitrates from 16 to 128 kbps/channel [7]. Speex [8] is based on the Code Excited Linear Prediction (CELP) [9] algorithm and, unlike the previously existing Vorbis codec, is optimised for transmitting speech for low latency communication over an unreliable packet network. Deep Neural Networks (DNN) [10], Convolutional Neural Networks (CNN) Variational AutoEncoder (AE) [11] along with Generative Adversarial Networks (GAN) [12] are among the important AI networks to apply to audio coding. AE networks by using an encoder-decoder framework with an information bottleneck in-between creates the possibility of designing low-bit-rate speech. WaveNet [13] is among the pioneer audio generative models based on PixelCNN architecture [14]. Rather than coding, WaveNet can provide a generic and flexible framework for tackling many other applications that rely on audio generation such as Text-To-Speech (TTS) [15], speech enhancement, [16], and voice conversion [17].

Since deep learning approaches could condition on certain features, the combination of conventional algorithmic approaches with new deep learning could result in low-bit rate codecs. Kleijn et al. [18] used a learned WaveNet decoder to produce audio from the bit stream generated by the encoder in Codec2 (a parametric codec designed for low bit rates) operating at 2.4 kbps. This demonstrates how a learned decoder can better reconstruct the original signal over conventional hand-engineered algorithmic decoders even by using a low-bit rate encoder [3]. Moreover, VQVAE [19] is a frame that can encode speech into a compact discrete latent representation and then reconstruct the original audio by a decoder conditioned on certain features. In fact, can be used as an end-to-end speech codec mapping the input audio to the discrete latent space learned by VQ-VAE and then reconstructing the signal using the decoder network conditioned on the latent representation [3].

A combination of VQVAE and WaveNet to create an architecture suited to the task of speech coding has been introduced in [3]. This architecture reduces 16-bit pulse-code modulation (PCM) encoded speech at 16 kHz sample rate (256 kbps) to a stream of discrete latent codes at 4 kbps. An important point about utilizing VQVAE in speech coding is to maintain input prosody at the output reconstructed signal. Gârbacea et al. suggests using a loss term to preserve pitch (f0) and timing information. Recently released, SoundStream [20] is an end-to-end neural audio codec that can efficiently compress speech, music, and general audio at normal bitrates. SoundStream relies on a model architecture composed of a fully convolutional encoder/decoder network and a residual vector quantizer, which are trained jointly end-to-end. SoundStream at 3 kbps outperforms Opus [21] at 12 kbps and approaches EVS [22] at 9.6 kbps. In section II, we first introduce the proposed approach and then in sections III and IV, the experimental setting and assessment results are explained.

## PROPOSED APPROACH

In this paper, we propose an AI audio coding approach based on adversarial learning. One of the problems of conventional coding is reconstruction artifacts. You may extract compact and informative representations of the signal but fail to reconstruct the original signal thoroughly. A good representation does not necessarily guarantee a good reconstruction. On the other hand, if enough data is available, AI approaches could learn the nonlinear transform better and therefore, compensate for the defects usually occur in the decoding process of conventional audio codecs. What makes AI better than conventional reconstruction approaches comes back to the fact that learning methods (e.g. AI) take into account both representing features as input and the original signal as target simultaneously at the training phase. Whereas, conventional methods only have access to the extracted features and not the original signal as a target. Therefore, AI-based approaches generally have a high capacity for reconstruction if big data is available.

In this paper, we have used Mel-Frequency Cepstral Coefficients (MFCCs) for our feature extraction. These features are typically used in speech recognition [23] and music information retrieval [24]. As explained, the essential problem of conventional modelling locates the reconstruction signal out of extracted features. To overcome this problem, we suggest using a learned decoder for reconstruction. We extract audio coefficients and use the adversarial loss to generate output raw waveform. Since Generative Adversarial Networks (GANs) produce explicit density, thus, we choose such architecture for our decoder framework.

## GENERATIVE ADVERSARIAL NETWORKS (GAN)

Generative Adversarial Network (GAN) was first introduced in 2014 by Ian Goodfellow [12]. Since GAN gives implicit density estimation, therefore, such models are less prone to overfitting. GANs consist of two integral parts: generator and discriminator.

The generator tries to convert the input into the desired output. On the other hand, the discriminator acts as a strict teacher to detect real and fake data (generator output). The generator tries to fool the discriminator by generating an output similar to real target data, and then again, the discriminator tries to improve itself and not be fooled. The interaction between the generator and the discriminator finally ends with a better signal reconstruction.

*Figure 1* shows the general structure of GAN systems operating on speech signals. We now describe the structure of the generator and discriminator adopted in our paper:

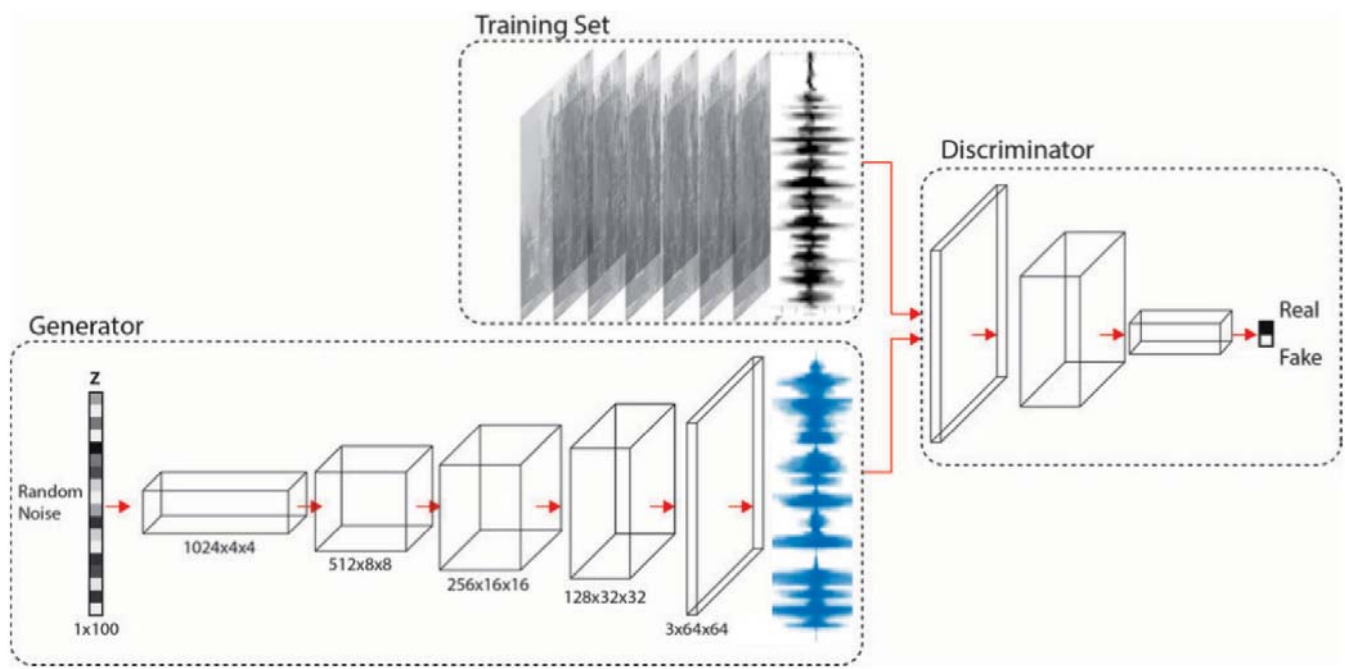

*Figure 1: General structure of GAN*

## GENERATOR

The generator adopted here consists of fully convolutional layers converting input Mel-spectrogram to output raw waveform [1]. The output is 256X higher temporal resolution compared to the input. To upsample the resolution, a stack of transposed convolutional layers is used in sequence. Besides, each convolution layer is followed by a residual block consisting of dilated convolutions. The entire structure of the generator is illustrated in *Figure 2(a)*.

## DISCRIMINATOR

Discriminator consists of a multi-scale architecture with three discriminators (D1, D2, D3) that have an identical network structure each operating at a different output audio scale [1]. Scaling output audio signals reveals different features; if high-frequency properties are required, a high temporal resolution signal should be taken into account. On the other hand, if low-frequency properties are desired, low-temporal resolution signals are considered. Finally, the use of three discriminators makes the final discriminator stricter and thus, forces the generator to generate signals more similar to the original one. It is illustrated in *Figure 2(b)*.

## LOSS FUNCTION

The objective functions used in the training setting are as follows [1]:

$$\min_{D_k} E_x\left[\min(0, 1-D_k(x)) + E_{s,z}[\min(0, 1+D_k(G(s,z)))]\right] \quad (1)$$
$$\forall k = 1,2,3$$

$$\min_{G} E_{s,z}\left[\sum_{k=1,2,3} -D_k(G(s,z))\right] \quad (2)$$

$$L_{FM}(G, D_k) = E_{x,s\sim p_{data}}\left[\sum_{i=1}^{T} \frac{1}{N_i} \left\| D_k^{(i)}(x) - D_k^{(i)}(G(s)) \right\|_1\right] \quad (3)$$

$$\min_{G}\left(E_{s,z}\left[\sum_{k=1,2,3} -D_k(G(s,z))\right] + \lambda \sum_{k=1}^{3} L_{FM}(G, D_k)\right) \quad (4)$$

Equation (1) and (2) tries to train the discriminator and the generator respectively. As explained, to increase the performance of the generator, the multi-scale discriminator adds LFM according to equation (3) and finally makes the total generator loss function according to (4).

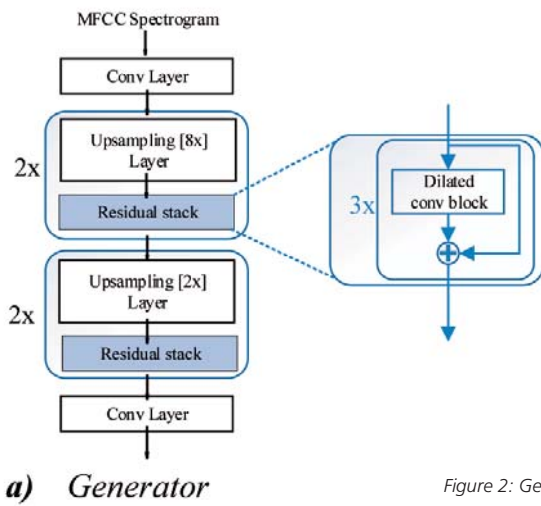
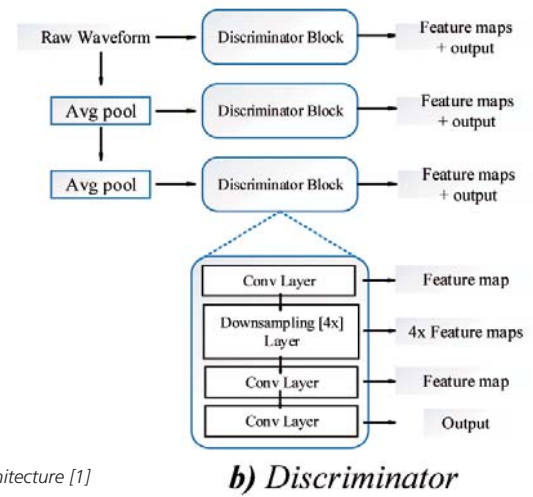

*Figure 2: Generator and Discriminator architecture [1]*

## EXPERIMENTAL SETTING

To experiment with the results, we extracted MFCC at each frame and fed the features to the network and trained the network to generate output raw waveform. For our experiment, 13, 24, and 36 coefficients were extracted and evaluated respectively. We utilised The LJSpeech Dataset [25] which consists of 13100 audio waveforms. We separated 13090 for our training set and 10 samples for our test set with no overlapped samples. The learning rate was set to 1e-4 for both the generator and discriminator. We trained our models for 3000 epochs.

## SUBJECTIVE AND OBJECTIVE RESULTS

We compared our model with five well-known codecs currently used in broadcasting and streaming technologies. The selected codecs cover bitrates ranging from 2kbps to 128kbps. We assessed our results using both objective and subjective assessments. We used SNR, STOI [26], PESQ [27], and NISQA-MOS [28] metrics to objectively evaluate our generated waveforms. To better discriminate between codecs, we choose <codec name_bitrate> format to name files and their corresponding bitrate simultaneously. According to table I, MFCCGAN_36k achieved the state-of-art result in terms of SNR despite a lower bitrate in comparison to AC3_128k, AAC_112k, Vorbis_48k, Opus_48k, and Speex_48K. On the other hand, MFCCGAN-13k also achieved high SNR=27 which is equal to that of AC3_128k, and AAC_112k while having a significantly lower bitrate (13 kbps). MFCCGAN_25k also gained SNR=25 which is close to AAC_112 and AAC_96k. SNR indicates the ability of the codec to reconstruct the original waveform.

MFCCGAN high SNRs reveal the capacity of AI approaches for reconstruction. While AI approaches act based on learning,

rule-based approaches utilise fixed algorithmic methods. Therefore, in the case of big data, AI could outperform conventional methods. NISQA-MOS represents subjective naturalness via objective assessment.

MFCCGAN_36k achieved higher NISQA-MOS results compared to AAC_48k while having a 20% lower bitrate. Furthermore, MFCCGAN_13k obtained NISQA-MOS=3.9 which is much higher than AAC_24k (NISQA-MOS=2.7), AAC_32k and AC3_32k (NISQA-MOS=2.6, 3.4 respectively) and AAC_48k (NISQA-MOS=3.05). STOI and PESQ illustrate the intelligibility and perceptual quality of speech. Accordingly, MFCCGAN is not as good as other codecs in terms of intelligibility and perceptual performance. We suggest adopting loss functions optimising on intelligibility and perceptual metrics. This will be taken into account in our future work.

## CONCLUSION AND FUTURE WORKS

In this paper, we proposed AI-based audio coding using MFCC features in an adversarial setting. Based on the application and options adopted, bitrates ranging from 13kbps to 64kbps could be achieved. We compared our work with five well-known codecs named AAC, AC3, Opus, Vorbis, and Speex, performing on bitrates from 2kbps to 128kbps. MFCCGAN_36k achieved the state-of-art result in terms of SNR despite a lower bitrate in comparison to AC3_128k, AAC_112k, Vorbis_48k, Opus_48k, and Speex_48K. On the other hand, MFCCGAN_13k also achieved high SNR=27 which is equal to that of AC3_128k, and AAC_112k while having a significantly lower bitrate (13 kbps). MFCCGAN_36k achieved higher NISQA-MOS results compared to AAC_48k while having a 20% lower bitrate. Furthermore, MFCCGAN_13k obtained NISQA-MOS=3.9 which is much higher than AAC_24k, AAC_32k and AC3_32k, and AAC_48k.

For future work, we suggest adopting loss functions optimising on intelligibility and perceptual metrics in the MFCCGAN structure.

*Table 1: Objective assessment of codecs*

| CODEC_BitRate | STOI | PESQ | SNR | NISQA-MOS |
|---|---|---|---|---|
| AC3_128K | 0.966639246 | 3.842376232 | 13.99329433 | 4.512829 |
| AAC_112K | 0.991996208 | 3.583273888 | 27.94281862 | 4.39592 |
| AAC_96K | 0.989615808 | 3.614152193 | 27.50264894 | 4.38134 |
| AC3_64K | 0.956902768 | 3.772472858 | 14.07268355 | 4.346048 |
| VORBIS_48K | 0.982319119 | 4.315928459 | 19.89123884 | 4.473292 |
| AAC_48K | 0.970645834 | 3.467241287 | 22.44192965 | 3.051058 |
| OPUS_48K | 0.929507751 | 3.851855278 | 14.00275241 | 4.478506 |
| SPEEX_48K | 0.90961242 | 3.55660224 | 14.78445809 | 4.278224 |
| AAC_40K | 0.959075887 | 3.39499712 | 21.35531287 | 2.968935 |
| MFCCGAN_36K | 0.702409688 | 3.093079567 | 30.83065793 | 3.995804 |
| AAC_32K | 0.953966379 | 3.297092915 | 19.61904286 | 2.610587 |
| AC3_32K | 0.89188321 | 2.463285685 | 10.81502084 | 3.462113 |
| MFCCGAN_25K | 0.598285956 | 1.801628828 | 25.97138144 | 2.55628 |
| SPEEX_24K | 0.877076344 | 3.000904083 | 12.91462434 | 3.603933 |
| VORBIS_24K | 0.94226169 | 3.515651703 | 14.62278805 | 3.006679 |
| OPUS_24K | 0.92259971 | 3.945333719 | 13.6733529 | 4.448306 |
| AAC_24K | 0.884156106 | 2.425422668 | 14.82544485 | 2.711517 |
| VORBIS_16K | 0.922521178 | 3.019296646 | 13.40416934 | 2.279955 |
| OPUS_16K | 0.889483445 | 3.670092344 | 12.87143345 | 4.375881 |
| SPEEX_16K | 0.831684321 | 2.83521533 | 12.63241821 | 3.623759 |
| AAC_16K | 0.790138556 | 1.741425514 | 9.751237205 | 3.149007 |
| MFCCGAN_13K | 0.447201272 | 1.155933857 | 27.21439393 | 3.965796 |
| SPEEX_12K | 0.819116709 | 2.77005887 | 12.03720997 | 3.418867 |
| OPUS_12K | 0.886875364 | 3.037414312 | 12.68146817 | 3.768796 |
| SPEEX_10K | 0.819116709 | 2.77005887 | 12.03720997 | 3.418867 |
| AAC_8K | 0.60164674 | 1.568306923 | 8.883444036 | 1.54391 |
| VORBIS_8K | 0.922521178 | 3.019296646 | 13.40416934 | 2.279955 |
| OPUS_8K | 0.804094524 | 1.997177839 | 9.142479641 | 2.937732 |
| OPUS_6K | 0.763997101 | 1.546600819 | 7.389424542 | 2.763465 |
| SPEEX_5K | 0.665002901 | 1.616167545 | 6.97172723 | 1.809035 |
| SPEEX_2K | 0.665002901 | 1.616167545 | 6.97172723 | 1.809035 |

## AUTHOR

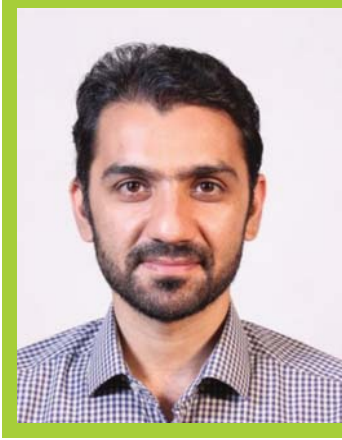

### Mohammad Reza Hasanabadi
*IRIB R&D, Iran*

Mohammad Reza Hasanabadi received his B.S. in electrical engineering from Ferdowsi University of Mashhad in 2013, and his M.S. in Sound Engineering from IRIB University in 2017. He is currently a Ph.D. candidate in telecommunication engineering at Shahid Beheshti University (SBU), Tehran, IRAN. In 2019, he joined IRIB R&D. His research interests include deep learning and AI audio video fields particularly focusing on speech processing, voice conversion, text-to-speech, audio coding, and their applications to broadcast systems.